\documentclass[12pt]{article}
\usepackage{amsfonts}
\usepackage{amssymb}
\usepackage{amsmath}
\usepackage[T2A]{fontenc}
\usepackage[cp1251]{inputenc}
\numberwithin{equation}{section}
\renewcommand{\Re}{\text{Re}}\renewcommand{\Im}{\text{Im}}

\begin{document}

\title{Equation with the lower negative time number in the Davey--Stewartson hierarchy.}
\author{A.~K.~Pogrebkov\\
Steklov Mathematical Institute of\\ Russian Academy of Sciences;\\
HSE University, Department of Mathematics;\\
Krichever Center for Advanced Studies at Skoltech;\\ 
Moscowof times; hierarchies; dimensional reductions.} 

\maketitle

\begin{abstract}
In \textit{SIGMA} \textbf{17} (2021), 091, 12 p.p.\ we have presented an integrable system with a negative time variable number for the Davey-Stewartson hierarchy. Here we develop this approach to construct an integrable equation with a lower time variable number. In addition, we show that the system reduced by this time is a new integrable equation in the dimension $1+1$.
\end{abstract}
 
\section{Introduction}
Problems with negative symmetries, i.e., symmetries with negative numbers, lead to the construction of new classes of integrable systems, in particular in the $2+1$ dimension. The initial ideas of constructing $2+1$ dimensional integrable systems go back to the works of \cite{ZShabat}, \cite{ZM} and \cite{ZShu}. Thus, in \cite{ZShu} it was proved that a $2+1$ dimensional dynamical system is integrable in the case of a degenerate conservation law. In \cite{akp2008a}, this approach was generalized through the use of commutator identities, the essence of which boils down to the following. Let $A$, $B$, and $\sigma$ be elements of the associative algebra $\mathcal{A}$ with unit $I$, such that $\sigma^2=I$, $[\sigma,A]=0$, and $\{\sigma,B\}=$0. Here $\{\cdot,\cdot\}$ denotes an anticommutator, so $[\sigma A,B]=\sigma\{A,B\}$. Then we have two commutator identities:
\begin{equation}
\sigma[A^2,B]=[A,[\sigma A,B]],\quad\sigma[\sigma A^2,B]=[A,[A,B]]+
[\sigma A,[\sigma A,B]],\label{f1}
\end{equation}
which relate to the derivatives $A$ and $B$. Introducing the dependence of the operator $B$ on two sets of times $t=\{t_1,t_2, t_3,\ldots\}$ and $x=\{x_1,x_2,x_3,\ldots\}$ by means of the relations 
\begin{equation}
B_{t_n}=[A^n,B],\qquad B_{x_n}=[\sigma A^n,B],\label{f2}
\end{equation}
we write (\ref{f1}) in the form of two linear differential equations:
\begin{equation}
\sigma B_{t_2}=B_{t_1x_1}\text{and } \sigma 
B_{x_2}=B_{t_1t_1}+B_{x_1x_1},\label{f3}
\end{equation}
so that each of them gives a linearized versions of the Davey--Stewartson equation (DS). 

This approach was developed in application to many integrable differential, differential-difference and difference equations in $2+1$, where it was applied to the case of positive numbers of $n$ in relations (\ref{f2}), i.e., to positive time numbers. In \cite{akp2021a} and \cite{akp2021b}, we generalized this approach to the case where these numbers can be negative. In particular, we considered the case based on the commutators $[A,B]$, $[\sigma A,B]$, and $[A^{-1},B]$, or $[A,B]$, $[\sigma A,B]$, and $[\sigma A^{-1},B]$, where we assumed the existence of the inverse element $A^{-1}$. It is easy to verify that these commutators satisfy the following commutator identities:
\begin{align}
&[\sigma A,[\sigma A,[A^{-1},B]]]-[A,[A,[A^{-1},B]]]+4[A,B]=0,\label{commut3}\\
&[\sigma A,[\sigma A,[\sigma A^{-1},B]]]-[A,[A,[\sigma A^{-1},B]]]-4[\sigma A,B]=0.\label{commut4}
\end{align}
Taking into account that all these commutators mutually commute, we consider $B$ as a function of $t_1$, $x_1$ and $t_{-1}$, or $x_{-1}$ such that
\begin{subequations}\label{comm2}
\begin{equation}
 B_{t_1}=[A,B],\qquad B_{x_1}=[\sigma A,B],\label{comm21}
\end{equation}
 \begin{equation}
B_{t_{-1}}=[A^{-1},B],\qquad B_{x_{-1}}=[\sigma A^{-1},B],\label{comm22}
\end{equation}
\end{subequations}
that  extends (\ref{f2}) to the negative values of $n$. Under these definitions Eqs.\ (\ref{commut3}) and (\ref{commut4})
prove that $B(t,x)$ obeys the following linear equations of motion:
\begin{align}
&B_{x_1x_1t_{-1}}-B_{t_1t_1t_{-1}}+4B_{t_1}=0,\label{ci1}\\
&B_{x_1x_1x_{-1}}-B_{t_1t_1x_{-1}}-4B_{x_1}=0.\label{ci2}
\end{align}
In \cite{akp2021a} and \cite{akp2021b} we derived dressing procedure that enabled nonlinearization of these equations, i.e., constraction of Lax pairs and integrable nonlinear equations. In the case where the set of independent variables includes $t_1$, $x_1$, and $t_{-1}$ the nonlinear evolution equation sounds as
\begin{equation}
u_{t_1t_{-1}}\sigma-u_{x_1t_{-1}}-[\sigma,\psi(1+u_{t_{-1}})]+[u_{t_{-1}},[\sigma,u]]=0.\label{eq1} 
\end{equation}
If this set includes $t_1$, $x_1$, and $x_{-1}$, then the nonlinear integrable equation has the form
\begin{equation}
u_{t_1x_{-1}}\sigma-u_{x_1x_{-1}}-[\sigma,\psi(\sigma+u_{x_{-1}})]+[\sigma+u_{x_{-1}},[\sigma,u]]=0.\label{eq3}
\end{equation}
Here $\psi$ is an auxiliary variable defined by the constraint
\begin{equation}
\psi_{x_1}-\sigma\psi_{t_1}-[\sigma,u_{t_1}]+[\sigma,\psi]\psi+[[\sigma,u],\psi]=0,\label{eq2} 
\end{equation}
the same for both equations (\ref{eq1}) and (\ref{eq2}), $u$ and $\psi$ are $2\times2$ matrices and 
\begin{equation}
\sigma=\sigma_3.\label{sigma}                                                                                                                           \end{equation}

Equations in $2+1$ that involve times with positive numbers only admit reductions to $1+1$ integrable nonlinear equations, say, reduction of the DS equation to Nonlinear Schr\"odinger equation. But equations (\ref{eq1})--(\ref{eq3}) under reduction $u_{t_{-1}}=0$ (or $u_{x_{-1}}=0$) reduce to the identity $0=0$.

In this article we consider the lower equation of this hierarchy, i.e., equation defined by commutators (\ref{comm21}) and 
\begin{equation}
B_{x_{-2}}=[\sigma A^{-2},B].\label{f4} 
\end{equation}
We omitt here equation that follows from the choice of coordinates (\ref{comm21}) and $B_{t_{-2}}=[A^2,B]$, because its construction is pretty close to this one, as follows from the above. Commutators $[A,B]$, $[\sigma A,B]$, and $[\sigma A^2,B]$ obey identity
\begin{align}
&[\sigma A^{-2},\underbrace{[\sigma A,\ldots,[\sigma A}_{4},B]\ldots]- 2[\sigma A,[\sigma A,[A,[A,B]]]]+\nonumber\\
+&\underbrace{[A,\ldots,[A}_{4},B]\ldots]]=
8\sigma(\bigl[\sigma A,[\sigma A,B]]+[A,[A,B]]\bigr),\label{f5}
\end{align}
that thanks to (\ref{comm21}) and (\ref{f4}) gives that $B$ obeys the following linear differential equation
\begin{equation}
\partial_{x_{-2}}^{}\bigl(\partial_{x_1}^{2}-\partial_{t_1}^2\bigr)^2B=
8\sigma\bigl(\partial_{x_1}^2+\partial_{t_1}^2\bigr)B. \label{f6}
\end{equation}

In the next section we present dressing procedure that enables derivation of the Lax pair by means of the $\overline{\partial}$ problem for the dressing operator. In Sec.\ 3 we construct the Lax pair and show that for completeness we need an intermidiate step: discrete transformation. Equation of compatibility, i.e., system of nonlinear evolution equations is given in 
Sec.\ 4. Then, in Sec.\ 5 we consider $1+1$ dimensional reduction of the constructed system. Some concluding remarks are given in Sec.\ 6.

\section{Dressing procedure.}
Following \cite{akp2008a}, \cite{akp2021a} we consider a set of pseudo-differential operators with respect to variable {$t_1$} with symbols depending on real variables $x_1$, $x_{-2}$, and complex variable $z\in\mathbb{C}$.  
These operators, denoted as $F(t_1,x,z)$, $G(t_1,x,z)$, etc., where $x=\{x_1,x_{-1}\}$, obey a composition law
\begin{equation}
(FG)(t_1,x,z)=\dfrac{1}{2\pi}\int dp\int dy_1\,
F(t_1,x,z+ip)e^{ip(t_1-y_1)}G(y_{1},x,z).\label{f7}
\end{equation}
We consider only those operators symbols of which belong to the space of tempered distributions. In this space of pseudo-differential operators we realize operators $A=A(t_1,x,z)$, $B=B(t_1,x,z)$, etc. Then we define
\begin{equation}
A(t_1,x,z)=z,\label{f8}                                                                                                                                                                                                                                       
\end{equation}
so that due to (\ref{f7}) for an arbitrary pseudo-differential operator $F(t_1,x,z)$ we get
\begin{equation}
(AF)(t_1,x,z)=(\partial_{t_1}+z)F(t_1,x,z),\qquad (FA)(t_1,x,z)=zF(t_1,x,z).
\label{f9}
\end{equation}
Thus $[A,F]=\partial_{t_1}F$, as it must be due to (\ref{comm21}).

Moreover, for any {$F$} and any $m\in\mathbb{Z}$ we have also:
\begin{align}
(A^mF)(t_1,x,z)&=(z+\partial_{t_1})^{m}F(t_1,x,z),\quad m\geq0,\label{f10}\\
(A^mF)(t_1,x,z)&=\dfrac{\text{sgn}(z_{\Re})}{(-m-1)!}\int dy_1\,y_{1}^{-m-1}
e^{-y^{}_1z}\theta(y_1z_{\Re})\times\nonumber\\
&\times F(t_1-y_1,x,z),\quad m<0,\label{f11}\\
(FA^m)(t_1,x,z)&=z^{m}{F}(t_1,x,z),\label{f12}
\end{align}
where $z\in\mathbb{C}$ and where $A^m$ is understood as $m$th power of composition (\ref{f7}), $\theta(x_1)$ denotes Heaviside step-function. Thus composition law (\ref{f7}) results in the normal ordering of the product of pseudo-differential operators, where all operators $A$ are shifted to the right by means of (\ref{f9}).

Because of our definition, the symbol $B(t,x,z)$ of operator $B$ admits a Fourier transform with respect to the variable $t_1$, so one can solve relations (\ref{comm21}) and (\ref{f7}) as
\begin{align}
B(t_1,x,z)=&\int dp\,\exp\bigl(ipt_1+\sigma(2z+ip)x_1+
\nonumber\\
&+\sigma\bigl(1/(z+ip)^2+1/z^2\bigr)x_{-1}\bigr)f(p,z),\label{Bf} 
\end{align}
where $f(p,z)$ is an arbitrary $2\times2$ antidiagonal matrix function independent of $t_1$ and $x$. It is natural to impose on $B(t_1,x,z)$ conditions of convergence of the integral and the boundedness of the limits of $B(t_1,x,z)$ as $t_1$, or $x$ tend to infinity. The choice sufficient for this condition is given by $f(p,z)=\gamma(p+2z_{\Im})g(z)$ so that (\ref{Bf}) takes the form
\begin{equation}
B(t_1,x,z)=\exp\biggl(\bigl(\overline{z}-z\bigr)t_1+\sigma\bigl(\overline{z}+z\bigr)x_1+\sigma\bigl(1/\overline{z}^2+1/z^2\bigr)x_{-1}\biggr)g(z),\label{Bf1} 
\end{equation}
where $g(z)$ is an arbitrary bounded function of its argument. In order to get $B(t_1,x,z)$ bounded with respect to variables $x_1$ and $x_2$ we have to substitut
\begin{equation}
x_1\to ix_1,\qquad x_{-1}\to ix_{-1},\label{bf1}
\end{equation}
with real $x_n$'s.

Another choice sufficient for boundedness of $B(t_1,x,z)$ in (\ref{Bf}) is  $f(p,z)=\gamma(z_{\Re})h(p,z_{\Im})$. We omitt this case here as construction is similar  to the above, with the only difference that in this case we do not need substitution (\ref{bf1}). Existence of these two types of systems is characteristic for $2+1$ case (cf.\ KPI, KPII and DSI, DSII equations) that corresponds to the Inverse problems given either by $\overline\partial$-problem, or by nonlocal Riemann--Hilbert problem.

Specific property of this set of pseudo differential operators is possibility to define operation of $\bar\partial$-differentiation with respect to $z$:
\begin{equation}
F\to\bar\partial{F}:\qquad
(\widetilde{\bar{\partial}F})(t,z)=
\dfrac{\partial {\widetilde{F}(t,z)}}{\partial\overline{z}},\label{f13}
\end{equation}
where derivative is understood in the sense of distributions. In particular, thanks to (\ref{f8})
\begin{equation}
\bar{\partial}A=0.\label{A0}                    
\end{equation}
 
This definition enables introduction of the \textbf{dressing operator} $K$ as solution of the $\overline\partial$-problem
\begin{align}
&\overline\partial K=KB,\label{f14}\\
&K(t_1,x,z)\to I,\quad z\to\infty,\label{f15}
\end{align}
where $K(t_1,x,z)$ denotes symbol of this operator, $I$ is the $2\times2$ unity matrix and product in the r.h.s.\ of (\ref{f14}) is understood in the sense of composition law (\ref{f7}). Thanks to (\ref{Bf1}) Eq.\ (\ref{f14}) can be written explicitly as
\begin{align}
&\dfrac{\partial{K}(t_1,x,z)}{\partial\overline z}=
K(t_1,x,\overline{z})\times\nonumber\\
&\times\exp\biggl(\bigl(\overline{z}-z\bigr)t_1+\sigma\bigl(\overline{z}+z\bigr)x_1+\sigma\bigl(1/\overline{z}^2+1/z^2\bigr)x_{-1}\biggr)g(z).\label{f16} 
\end{align}

Evolution with respect to times $t_1$, $x_1$, and $x_{-2}$  follows from (\ref{f14}):
\begin{equation}
\overline\partial K_{t_1}=K_{t_1}B+K[A,B],\qquad \overline\partial K_{x_j}=K_{x_j}B+K[\sigma A^{j},B],\quad j=1,-2.\label{K-t}
\end{equation}
We impose condition of unique solvability of the problem (\ref{f14}), (\ref{f15}). Thanks to this condition we prove commutativity of the derivatives of the dressing operator with respect to independent variables  $t_1$, $x_1$, and $x_{-2}$. Say, differentiating (\ref{K-t}) we get 
\begin{align*}
\overline\partial K_{x_1x_{-2}}&=K_{x_1x_{-2}}B+K_{x_1}[\sigma A^{-2},B]+
K_{x_{-2}}[\sigma A,B]+K[\sigma A^{-2},[\sigma A,B]],\\
\overline\partial K_{x_{-2}x_1}&=K_{x_{-2}x_1}B+K_{x_1}[\sigma A^{-2},B]+
K_{x_{-2}}[\sigma A,B]+K[\sigma A,[\sigma A^{-2},B]], 
\end{align*}
so that taking commutativity of commutators in (\ref{comm21}) and (\ref{f4}) into account we get  
$\overline\partial(K_{x_1x_{-2}}-K_{x_{-2}x_1})=
(K_{x_1x_{-2}}-K_{x_{-2}x_1})B$. 
Thus commutativity of derivatives
\begin{equation}
K_{x_1x_{-2}}-K_{x_{-2}x_1}=0\label{K-tt} 
\end{equation}
in this approach follows directly from condition of the unique solvability of the Inverse problem (\ref{f14}), (\ref{f15}). 

\section{Time evolutions of the dressing operator.}
In order to derive evolution equations for the dressing operator $K$ defined in  (\ref{f14}), (\ref{f15}) we assume accuracy of the asymptotic expansion
\begin{equation}
K(t_1,x,z)=I+u(t_1,x)z^{-1}+v(t_1,x)z^{-2}+w(t_1,x)z^{-3}+o(z^{-3}),
\label{Kuvw} 
\end{equation}
where $u$, $v$, and $w$ are multiplication operators with respect to composition law (\ref{f7}), i.e., their symbols are independent of $z$. Thanks to this assumption we get from the Inverse problem, that 
\begin{equation}
K_{t_1}=[A,K],\label{t1}
\end{equation}
and
\begin{equation}
K_{x_1}+K\sigma A=\sigma AK-[\sigma,u]K,\label{x1}
\end{equation}
as follows from \cite{akp2008a}, \cite{akp2021a}. 

Let us consider evolution with respect to $x_{-2}$. Thanks to (\ref{K-t}) for $j=-2$, we get $\overline\partial K_{x_{-2}}=
K_{x_{-2}}B+KA^{-2}B-KBA^{-2}$. Taking into account that $\overline\partial$-derivative of $z^n$ equals to zero only in the case of $n\geq0$ (cf.\ (\ref{f8}) and (\ref{A0})) we have to multiply this equality by $A^2$ from the right:
\begin{equation}
\overline{\partial}\bigl(K_{x_{-2}}A^2+K\sigma\bigr)=\bigl(K_{x_{-2}}A^2+K\sigma\bigr)A^{-2}BA^2.\label{K}
\end{equation}
Thus situation where times has negative numbers is essentially different with respect to the positive numbers. We got the quantity $K_{x_{-2}}A^2+K\sigma$ that obeys $\overline{\partial}$-equation (\ref{f14}) but with substituted scattering data: $B\to A^{-2}BA^2$.

In \cite{akp2021b} we suggested to use an additional discrete evolution, or symmetry to resolve this problem. Indeed, following \cite{akp2010} we know that discrete evolutions are given by commutator identities for commutators in group sense (in contrast to commutators in the algebraic sense, (\ref{comm21}), (\ref{comm22}), (\ref{f4}), etc.). So we introduce transformation $ABA^{-1}$ and introduce new discrete variable $n\in\mathbb{Z}$ in $B$, besides variables $t_1$, $x_1$, and $z$, such that 
\begin{equation}
B^{(m)}=A^mBA^{-m},\quad m\geq0, \label{B1}
\end{equation}
where $B^{(m)}(t_1,x,n,z)=B(t_1,x,n+m,z)$, $K^{(m)}(t_1,x,n,z)=K(t_1,x,n+m,z)$. It is easy to check that this similarity transformation commute with times $t_1$, $x_1$, and $x_{-2}$, say: 
$(B^{(1)})_{x_{-2}}=(B_{x_{-2}})^{(1)}$. It also can be included in a commutator identity, say, for $m=2$ for (\ref{B1})
\begin{equation}
[A^2,B+A^2BA^{-2}]+\sigma[\sigma A^2,B-A^2BA^{-2}]=0,\label{f17} 
\end{equation}
so that notations (\ref{comm21}), (\ref{f4}), and (\ref{B1}) prove that $B$ obeys differential-difference equation
\begin{equation}
(B+B^{(2)})_{t_2}+\sigma(B-B^{(2)})_{x_2}=0.\label{f18} 
\end{equation}

Extending pointwise definition of composition law (\ref{f7}) to operators depending on $n$, we get 
\begin{equation}
\overline\partial K^{(2)}=K^{(2)}A^2BA^{-2}.\label{f19}                                           \end{equation}
At the same time, shit $n\to n+2$ in (\ref{K}) gives
\begin{equation}
\overline\partial\bigl(K^{(2)}_{x_{-2}}A^2+K^{(2)}\sigma\bigr)=\bigl(K^{(2)}_{x_{-2}}A^2+K^{(2)}\sigma\bigr)B,\label{K2}
\end{equation}
where (\ref{B1}) was used. Now, thanks to the Inverse problem  (\ref{f14}), (\ref{f15}), assumption of its unique solvability and equality (\ref{A0}) we prove existence of such multiplication operators $\widetilde\alpha$ and $\alpha$, that
\begin{equation}
K^{(2)}_{x_{-2}}A^2+K^{(2)}\sigma=\widetilde{\alpha} AK+\alpha K.\label{f20} 
\end{equation}
As in \cite{akp2021b} we get by (\ref{f15}) and decomposition (\ref{Kuvw}) that
\begin{equation}
\widetilde{\alpha}=u^{(2)}_{x_{-2}},\qquad \alpha=v^{(2)}_{x_{-2}}+\sigma -u^{(2)}_{x_{-2}}u.\label{f21} 
\end{equation}
In analogy, (\ref{f19}) demostrates that $\overline\partial \bigl(K^{(2)}A^{2}\bigr)=K^{(2)}A^2B$ due to (\ref{A0}), so that due to the inverse problem there exist multiplication operators $\beta$ and $\gamma$ such, that
\begin{equation}
K^{(2)}A^2=\bigl(A^2+\beta A+\gamma\bigr)K.\label{f22} 
\end{equation}
And again thanks to decomposition (\ref{Kuvw}) we derive, that
\begin{equation}
\beta=u^{(2)}-u,\qquad \gamma=v^{(2)}-v-2u_{t_1}-(u^{(2)}-u)u.\label{f23} 
\end{equation}

Inserting $K^{(2)}$ from (\ref{f22}) into (\ref{f20}) we derive thanks to  (\ref{f21}) and (\ref{f23}) 
\begin{align*}
&(A^2+\beta A+\gamma)(K_{x_{-2}}+KA^{-2}\sigma)-\\
-&u_{x_{-2}}AK-(2u_{t_1x_{-2}}+v_{x_{-2}}-u^{2}+\sigma)K=0.
\end{align*}
Taking that by (\ref{t1}) $AK=K_{t_1}+KA$ into account we get
equation
\begin{align}
&\bigl(K_{x_{-2}t_1t_1}+2K_{x_{-2}t_1}A+K_{t_1t_1}A^{-2}\sigma+
K_{x_{-2}}A^{2}+2K_{t_1}A^{-1}\sigma +K\sigma\bigr)+\nonumber\\
+&\beta\bigl(K_{x_{-2}t_1}+K_{x_{-2}}A+K_{t_1}A^{-2}\sigma+
KA^{-1}\sigma\bigr)-u_{x_{-2}}\bigl(K_{t_1}+KA\bigr)+\nonumber\\
+&\gamma\bigl(K_{x_{-2}}+KA^{-2}\sigma\bigr)+
(\gamma_{x_{-2}}-\alpha)K=0.\label{f24}
\end{align}

It looks that we constructed $(3+1)$-dimensional integrable system with independent variables $t_1$, $x_1$, $x_{-2}$, and $n$. But it is easy to see that while all these evolutions are mutually compatible, in fact we have a combination of three integrable systems with variables $t_1$, $x_1$, $n$ (positive numbers of times), $t_1$, $x_{-2}$, $n$ (version of the two dimensional Toda lattice) and $t_1$, $x_1$, $x_{-2}$, the latter because dependence on {$n$} can be excluded.

In analogy to the above construction we can use (\ref{t1}) to rewrite
(\ref{x1}) in the form
\begin{equation}
K_{x_1}+K\sigma A=\sigma K_{t_1}+\sigma KA-2[\sigma,u]K,\label{x}
\end{equation}
so that this equation and (\ref{f23}) are given now in the normally ordered form. This means that in terms of the symbols of operator $K$ and its derivatives we can substitute operator $A$ by the complex number $z$ due to (\ref{f12}). This enables introduction of of the Jost solutions instead of the dressing operator.

\section{Lax pair and equations of motion.}
We can simplify the above relations and to cancel dependence on operator $A$ by introducing the Jost solution, that in terms of the symbol of operator $K$ sounds as
\begin{equation}
\varphi(t_1,x,z)=K(t_1,x,z)e^{zt_1+\sigma zx_1+\sigma z^{-2}x_{-2}}.\label{f25}
\end{equation}
Then the Lax pair follows from Eqs.\ (\ref{x}) and (\ref{f24}):
\begin{align}
&\varphi_{x_1}-\sigma\varphi_{t_1}+[\sigma,u]\varphi=0,\label{f26}\\
&\varphi_{x_{-2}t_1t_1}+\beta\varphi_{x_{-2}t_1}-u_{x_{-2}}\varphi_{t_1}+
\gamma\varphi_{x_{-2}}+(\gamma_{x_{-2}}-\alpha)\varphi=0.\label{f27}
\end{align}
that results in the compatibility conditions:
\begin{align}
&\beta_{x_1}-\sigma\beta_{t_1}-[\sigma,\gamma]-2[\sigma,u_{t_1}]+
[\sigma,\beta]\beta+[[\sigma,u],\beta]=0,\label{f28}\\
&\gamma_{x_1}-\sigma\gamma_{t_1}-[\sigma,u_{t_1t_1}]+[[\sigma,u],\gamma]+[\sigma,\beta]\gamma-\beta[\sigma,u_{t_1}]=0,\label{f29}\\
&u_{x_1x_{-2}}-\sigma u_{t_1x_{-2}}+2[\sigma,u_{t_1x_{-2}}]-[\sigma,\alpha-\gamma_{x_{-2}}]+[[\sigma,u],u_{x_{-2}}]+\nonumber\\
&+[\sigma,\beta u_{x_{-2}}]=0,
\label{f30}\\
&\bigl(\alpha-\gamma_{x_{-2}}\bigr)_{x_1}-\sigma\bigl(\alpha-\gamma_{x_{-2}}\bigr)_{t_1}+
[\sigma,u_{t_1t_1x_{-2}}]+[\sigma,\beta]\bigl(\alpha-\gamma_{x_{-2}}\bigr)+\nonumber\\
&+[[\sigma,u],\alpha-\gamma_{x_{-2}}]+\beta[\sigma,u_{t_1x_{-2}}]-
u_{x_{-2}}[\sigma,u_{t_1}]+\gamma[\sigma,u_{x_{-2}}]=0.\label{f31}
\end{align}

Let us mention that the first two equations appear also as condition of compatibility of equations (\ref{f26}) and (\ref{f22}). In order to check this it is necessary to rewrite (\ref{f22}) by means of (\ref{f25}) as
\begin{equation}
\varphi^{(2)}z^2=\varphi_{t_1t_1}+\beta\varphi_{t_1}+\gamma\varphi.\label{f310}                                                                                                                                                                                 \end{equation}
This ``internal'' compatibility results in Eqs.\ (\ref{f28}) and (\ref{f29}), that here play the role of constraints: $\beta$ and $\gamma$ are defined by means of initial data $u(t_1,x_1,0)$. Eqs,\ (ref{f30}), (ref{f31}) also can be simplified. Taking  (\ref{f21}) and (\ref{f23}) into account we get that
\begin{equation}
\alpha-\gamma_{x_{-2}}=\sigma+v_{x_{-2}}+2u_{t_1x_{-2}}-u_{x_{-2}}u+\beta u_{x_{-2}},\label{f32}
\end{equation}
so this combination does not involve explicitly functions $u^{(2)}$ and $v^{(2)}$. Inserting this combination in Eq.\ (\ref{f30}), we write it in the form
\begin{equation}
\partial_{x_{-2}}\bigl(u_{x_1}-\sigma u_{t_1}-[\sigma,v]+[\sigma,u]u\bigr)=0.\label{f321} 
\end{equation}
This equality can be integrate with respect to $x_{-2}$. In what follows it is convenient to decompose $2\times2$ matrices $u$, $\beta$, and so on, into diagonal and anti-diagonal parts:
\begin{equation}
u=u^{\text{d}}+u^{\text{a}},\text{ etc}.,\label{f40}
\end{equation}
so that by (\ref{sigma}) $[\sigma,u^{\text{d}}]=0$, 
$[\sigma,u^{\text{a}}]=2\sigma u^{\text{a}}$, and so on. 
Then as a result we derive from (\ref{f321}) expression of the anti-diagonal part of matrix $v$:
\begin{equation}
[\sigma,v]=u_{x_1}-\sigma u_{t_1}+[\sigma,u]u=0,\label{f33} 
\end{equation}
and constraint for the diagonal part of matrix $u$:
\begin{equation}
u^{\text{d}}_{x_1}-\sigma u^{\text{d}}_{t_1}+2\sigma(u^{\text{a}})^2=0.\label{f34} 
\end{equation}
Now substitution of the r.h.s.\ of (\ref{f32}) in (\ref{f31}) gives dynamical equation for the diagonal part of matrix $u$ and anti-diagonal part of $v$.

\section{Dimensional reduction}
Reductions to $1+1$ dimensions follow from reductions of the linear equations (\ref{f6}). Taking solution of this equation in (\ref{Bf1}) we see that dependence on $t_1$ can be cancelled if $z_{\Im}=0$, dependence on $x_1$ if $z_{\Re}=0$, and dependence on $x_{-2}$ if $\overline{z}^2+z^2=0$. Also some linear combinations of these conditions can be considered. Thanks to the Inverse problem (\ref{f14}), (\ref{f15}) it is clear that any such reduction gives the corresponding reduction of independent variables of the dressing operator $K$. Here we consider only reduction with respect to the time $x_{-2}$:
\begin{equation}
z_{\Im}=\mp z_{\Re}.\label{f35}
\end{equation}
Thus Eq.\ (\ref{Bf1}) sounds now as
\begin{align}
B(t_1,x_1,z)&=e^{2z_{\Im}(it_1+\sigma x_1)}g_1(z_{\Im})\delta(z_{\Re}-z_{\Im})+\nonumber\\
&+e^{2z_{\Im}(it_1-\sigma x_1)}g_2(z_{\Im})\delta(z_{\Re}+z_{\Im}),\label{Bf2} 
\end{align}
where $g_1(z_{\Im})$ and $g_2(z_{\Im})$ are arbitrary bounded functions of their argument and we have to substitute $x_1\to ix_1$ according to (\ref{bf1}) in order to guarantee boundedness. Due to delta-functions in (\ref{Bf2}) the $\overline\partial$-problem (\ref{f14}) is substituted by the Riemann--Hilbert problem with discontinuities on the two straight lines in (\ref{f35}). We omit corresponding study for a forthcoming work and give here Lax pair and nonlinear equation.

Now the dressing operator $K$ is independent of $x_{-2}$ due to (\ref{f24}) and (\ref{x}), so we have to change definition (\ref{f25}) of the Jost solution: 
\begin{equation}
\phi(t_1,x_1,z)=K(t_1,x_1,z)e^{z(t_1+\sigma x_1)},\label{f36}
\end{equation}
so that Eqs.\  (\ref{f24}) and (\ref{x}) reduces to the following Lax pair:
\begin{align}
&\phi_{t_1t_1}+\beta \phi_{t_1}+\gamma\phi-z^{2}\sigma\phi\sigma=0,\label{f37}\\
&\phi_{x_1}=\sigma \phi_{t_1}-2[\sigma,u]\phi,\label{f38}
\end{align}
where we used that by (\ref{f21}) now $\alpha=\sigma$, while $\beta$ and $\gamma$ are still given by (\ref{f23}), but $u_{x_{-2}}=\gamma_{x_{-2}}=0$.

Compatibility condition of (\ref{f37}), (\ref{f38}) follows also as a reduction of Eqs.\  (\ref{f29})--(\ref{f31}), so that 
Eqs.\  (\ref{f28}) and (\ref{f29}) preserve their form, Eq.\ (\ref{f30}) cancels out due to (\ref{f321}), and Eq.\ (\ref{f31}) takes the form
\begin{equation}
[\sigma,\beta]\sigma+[[\sigma,u],\sigma]=0.\label{f39}
\end{equation}
Considering diagonal and antidiagonal (see (\ref{f40})) parts of matrices we see that the diagonal part of (\ref{f39}) cansels out from this equation and its anti-diagonal part gives $\beta^{\text{a}}=-2u^{\text{a}}$, that thanks to (\ref{f23}) is reduced to
\begin{equation}
(u^{(2)})^{\text{a}}=-u^{\text{a}}.\label{f41}
\end{equation}
Next, the diagonal part of (\ref{f37}) gives due to the above
$\beta^{\text{d}}_{x_1}-\sigma\beta^{\text{d}}_{t_1}=0$, that can be integrated to
\begin{equation}
\beta^{\text{d}}=0.\label{f42}
\end{equation}
Then anti-diagonal part of this equation defines anti-diagonal part of $\gamma$:
\begin{equation}
\gamma^{\text{a}}=-\sigma u^{\text{a}}_{x_1}-u^{\text{a}}_{t_1}.\label{f43} 
\end{equation}
Combining above results (\ref{f41})--(\ref{f43}) we reduce the diagonal part of (\ref{f30}) to
\begin{equation}
\gamma^{\text{d}}_{x_1}-\sigma\gamma^{\text{d}}_{t_1}-
2(u^{\text{a}})^2_{x_1}-2\sigma(u^{\text{a}})^2_{t_1}=0.
\label{f44}
\end{equation}
Finally, the anti-diagonal part of (\ref{f30}) after substitution of these equalities reads as
\begin{equation}
u^{\text{a}}_{t_1t_1}+u^{\text{a}}_{x_1x_1}+2
\{u^{\text{a}},\gamma^{\text{d}}\}=0.\label{f45} 
\end{equation}
Thus we have integrable system in $1+1$ dimensions, given by equations (\ref{f44}) and (\ref{f45}).

\section{Conclusion}
We proved that times with  lower negative numbers lead to new integrable systems in $2+1$ dimensions. Constructions of these systems is rather straightforward and follows the same ideas, as construction of integrable systems in the case of positive numbers of times. Essential modification of this approach is necessity to introduce some intermidiate variables, that play the role of constraints and that correspond to discrete evolutions.  

\section*{Conflicts of interest.} The author declares no conflicts of interest.

\end{document}